# Gaseous atomic nickel in the coma of interstellar comet 2I/Borisov

Piotr Guzik[1], Michał Drahus[1]

[1] Astronomical Observatory, Jagiellonian University, ul. Orla 171, 30-244 Kraków, Poland

**On 31 August 2019, an interstellar comet was discovered as it passed through the Solar System (2I/Borisov). Based on initial imaging observations, 2I/Borisov appeared to be completely similar to ordinary Solar System comets[1,2]—an unexpected characteristic after the multiple peculiarities of the only previous known interstellar visitor 1I/'Oumuamua[3-6]. Spectroscopic investigations of 2I/Borisov identified the familiar cometary emissions from CN (refs. [7-9]), $C_2$ (ref. [10]), O I (ref. [11]), $NH_2$ (ref. [12]), OH (ref. [13]), HCN (ref. [14]) and CO (ref. [14,15]), revealing a composition similar to that of carbon monoxide-rich Solar System comets. At temperatures >700 K, comets additionally show metallic vapors produced by the sublimation of metal-rich dust grains[16]. However, due to the high temperature needed, observation of gaseous metals has been limited to bright sunskirting and sungrazing comets[17-19] and giant star-plunging exocomets[20]. Here we report spectroscopic detection of atomic nickel vapor in the cold coma of 2I/Borisov observed at a heliocentric distance of 2.322 au—equivalent to an equilibrium temperature of 180 K. Nickel in 2I/Borisov seems to originate from a short-lived nickel-bearing molecule with a lifetime of $340^{+260}_{-200}$ s at 1 au and is produced at a rate of $0.9 \pm 0.3 \times 10^{22}$ atoms s$^{-1}$, or 0.002% relative to OH and 0.3% relative to CN. The detection of gas-phase nickel in the coma of 2I/Borisov is in line with the concurrent identification of this atom (as well as iron) in the cold comae of Solar System comets[21].**

We observed 2I/Borisov with the X-shooter spectrograph of the Very Large Telescope at the European Southern Observatory (ESO) on 28, 30 and 31 January 2020 UT. At the time of the observations, the mean heliocentric and geocentric distances of the comet were 2.322 and 2.064 au, respectively, and the mean heliocentric velocity was +19.16 km s$^{-1}$. X-shooter is a medium-spectral-resolution instrument that simultaneously covers the spectral range between 3,000 Å and 2.5 µm divided into three arms: UVB, VIS and NIR. However, for the present analysis we use only the UVB data. The UVB arm was configured to provide a spectral resolution of 4,100. The slit field of view was 1.3 × 10.88 arcsec, but the length was trimmed in the analysis to the inner 10.24 arcsec with low and uniform noise, corresponding to 1,950 × 15,300 km at the comet distance. For further details on the observations and data processing, please refer to the Methods section.

A portion of the spectrum presented in Fig. 1a,b shows nine emission lines between 3,375 and 3,625 Å that are not associated with any species routinely or less frequently detected in Solar System comets. The two brightest lines are clearly visible in co-added spectra (both processed and unprocessed) from each night, and are invisible in equivalent spectra of the sky background; this effectively eliminates the possibility of confusion with a background source or instrumental artefacts.

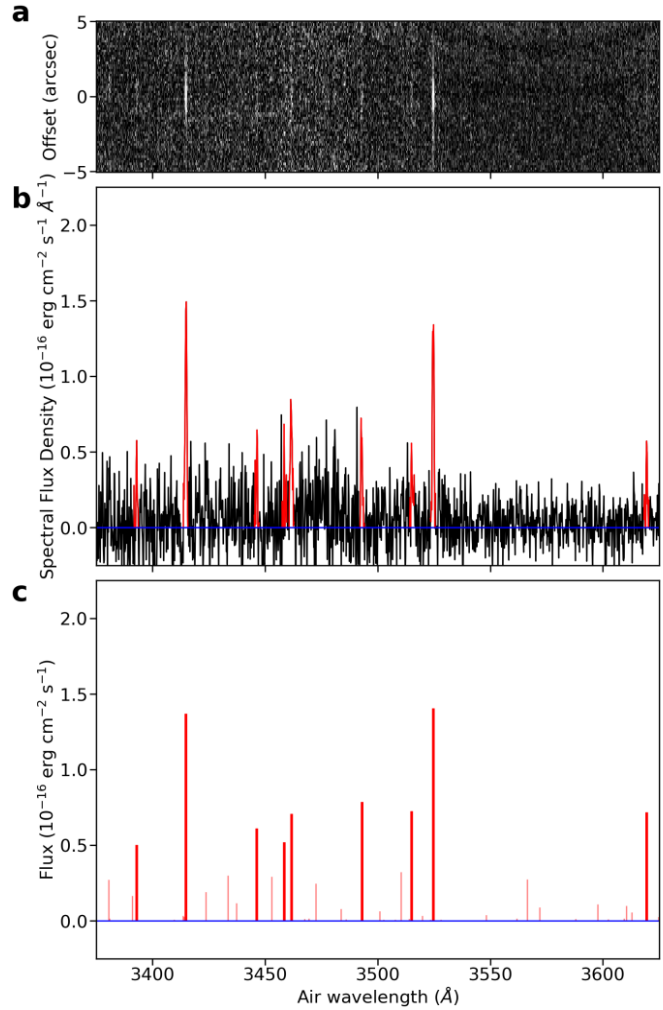

**Figure 1 | Emission lines from gaseous atomic nickel in the near-UV spectrum of 2I/Borisov. a,** Portion of the co-added and calibrated 2D spectrum with the dust-continuum component removed (see Methods). **b,** Corresponding 1D spectrum (see Methods) with the identified nickel emission lines highlighted in red. **c,** Modeled spectrum of nickel fluorescence emission (see Methods), converted to air wavelengths[22], and scaled to best match the two brightest lines.

Although the region between 3,375 and 3,625 Å does not overlap with the well known near-ultraviolet cometary emissions from OH, NH, and CN (all three also detected in our full spectrum from the UVB arm—see Extended Data Fig. 1), it does encompass the wavelengths of fainter emissions from OH, NH, and CN, as well as $H_2CO$, $S_2$, and the cometary ions $CO^+$, $CO_2^+$, $N_2^+$, and $OH^+$, which have all been previously detected or sought in this spectral region in other comets[16,23-27]. However, we find no combination of these species that would match the nine lines that we detected. Furthermore, from strong detections of OH at 309 nm, NH at 336 nm, and CN at 387 nm, and from non-detections of $CO^+$ at 425 and 427 nm, $CO_2^+$ at 367 nm, and $N_2^+$ at 391 nm (all wavelengths corresponding to the strongest bands of these species in our full spectrum), we conclude that these cometary species are undetectable in our data via the weaker emissions between 3,375 and 3,625 Å. We note that the detected nine lines are spatially compact (Fig. 1a) and spectrally unresolved (Fig. 1a,b), which renders their appearance distinct from that of the routinely observed compounds, such as OH, NH, and CN.

We recognize the detected lines as the spectroscopic signatures of atomic nickel vapor, Ni I, which has previously been observed in the spectrum of sungrazing comet C/1965 S1 (Ikeya-Seki)[17,18]. To confirm the identity, we created a model of nickel fluorescence emission (see Methods), which is dependent on the heliocentric distance and heliocentric velocity of a comet (the former scaling the amount of incoming solar energy through the inverse-square law and the latter arising as a result of the Swings effect[28]). A model spectrum computed for the epoch of our observations is presented in Fig. 1c. The agreement with the observed spectrum of 2I/Borisov is very good, although minor differences exist and can be attributed to several factors. On the observation side, these factors include the observation noise, minor features in the baseline, and rather large spectral channels compared to the line spread function (the last factor vanishing in the comparison of integrated line fluxes). The contributors on the model side are the limited accuracy of the input transition parameters, limited resolution, coverage and accuracy of the employed solar spectrum, and solar variability. Small differences are therefore not unexpected. The model predicts that no additional nickel lines will be detectable in other regions of the X-shooter spectrum (Extended Data Fig. 1). In Table 1 we show the measured fluxes of the detected nickel lines, their wavelengths and classifications, and the computed fluorescence efficiencies (see Methods).

**Table 1 | Measured, modeled and laboratory data of the detected nickel lines.**

| Laboratory air wavelength (Å) | Measured flux ($10^{-16}$ erg cm$^{-2}$ s$^{-1}$) | Modeled fluorescence efficiency at 1 au* ($10^{-13}$ erg s$^{-1}$) | Upper level | | | Lower level | | |
|---|---|---|---|---|---|---|---|---|
| | | | Conf. | Term | J | Conf. | Term | J |
| 3392.983 | 0.29 ± 0.12 | 0.6547 | $3d^9(^2D)4p$ | °  | 3 | $3d^9(^2D)4s$ | $^3D$ | 3 |
| 3414.764 | 1.44 ± 0.13 | 1.7834 | $3d^9(^2D)4p$ | $^3F°$ | 4 | $3d^9(^2D)4s$ | $^3D$ | 3 |
| 3446.259 | 0.48 ± 0.13 | 0.7963 | $3d^9(^2D)4p$ | $^3D°$ | 2 | $3d^9(^2D)4s$ | $^3D$ | 2 |
| 3458.460 | 0.42 ± 0.15 | 0.6772 | $3d^9(^2D)4p$ | $^3F°$ | 2 | $3d^9(^2D)4s$ | $^3D$ | 1 |
| 3461.652 | 0.88 ± 0.15 | 0.9196 | $3d^8(^3F)4s4p(^3P°)$ | $^5F°$ | 4 | $3d^9(^2D)4s$ | $^3D$ | 3 |
| 3492.956 | 0.56 ± 0.13 | 1.0223 | $3d^9(^2D)4p$ | $^3P°$ | 1 | $3d^9(^2D)4s$ | $^3D$ | 2 |
| 3515.052 | 0.51 ± 0.11 | 0.9423 | $3d^9(^2D)4p$ | $^3F°$ | 3 | $3d^9(^2D)4s$ | $^3D$ | 2 |
| 3524.536 | 1.34 ± 0.11 | 1.8276 | $3d^9(^2D)4p$ | $^3P°$ | 2 | $3d^9(^2D)4s$ | $^3D$ | 3 |
| 3619.391 | 0.44 ± 0.07 | 0.9349 | $3d^9(^2D)4p$ | $^1F°$ | 3 | $3d^9(^2D)4s$ | $^1D$ | 2 |

*Calculated for the heliocentric distance and heliocentric velocity of 2I/Borisov and reduced to 1 au from the Sun (see Methods).

The detected emission lines of nickel were further explored using the standard two-generation Haser model[29,30] (see Methods). First, we attempted to retrieve the parent and daughter scalelengths (see Methods) from the spatial profile of the lines along the slit, presented in Fig. 2. The sensitivity of spectral observations to these parameters is generally limited by the observation noise, and is further constrained by the spatial resolution (on the short side) and the extent of sampled coma region (on the long side). This restricts the usefulness of our data for such an analysis to the scalelengths (reduced to 1 au from the Sun) ranging from about one hundred kilometers up to a few tens of thousands kilometers. Figure 3a shows the results of the retrieval. It can be seen that the shorter scalelength reduced to 1 au from the Sun is found at 170 km, corresponding to 920 km at the heliocentric distance of comet 2I/Borisov. This value is comparable to—but slightly greater than—the radial extent of the point spread function (PSF) in our data (see Methods), corresponding to 750 km at the geocentric distance of the comet. The short scalelength is therefore reliably constrained by the data. Conversely, the long scalelength is found at the upper limit of the investigated space, implying that only its lower bound might be constrained by the data. To determine the allowable regimes of the two scalelengths, we ran a Monte Carlo simulation (see Methods), the result of which is presented in Fig. 3b. Within one standard error, we find the shorter scalelength to be confined to a region between 70 and 300 km and the longer scalelength to be >18,000 km, both reduced to 1 au. Assuming a constant gas expansion velocity of 0.5 km s$^{-1}$ (ref. [14]) and a negligible daughter excess speed, we find the corresponding lifetimes to be between 140 and 600 s and >36,000 s, respectively (also at 1 au). Although the Haser model alone does not enable us to distinguish which lifetime belongs to which generation, we can convincingly attribute the longer lifetime to nickel itself, consistent with its low photoionisation rate of $9.43 \times 10^{-7}$ s$^{-1}$ (ref. [32]), equivalent to a lifetime of $1.06 \times 10^{6}$ s (both at 1 au from the Sun).

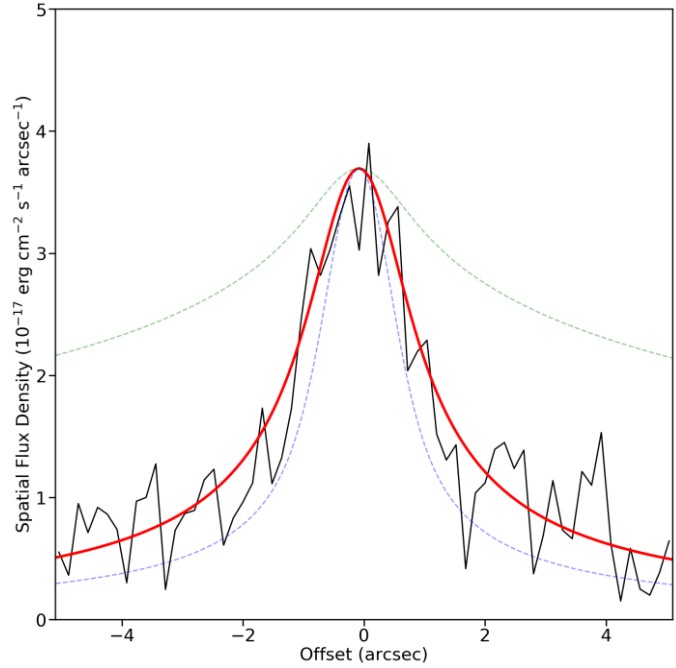

**Figure 2 | Observed and modeled spatial profiles of nickel emission.** The observed profile (black) is an average from the profiles of the two brightest lines. The modeled profile (red) was calculated using the standard two-generation Haser model with the best-matching combination of the parent and daughter scalelengths (see Methods). The standard deviation of the modeled profile from the observed profile is equal to $3.6 \times 10^{-18}$ erg s$^{-1}$ cm$^{-2}$ arcsec$^{-1}$. For reference, also shown is a two-generation Haser profile calculated with the canonical CN scalelengths[31] (green) and a single-generation profile calculated with an infinite scalelength (blue). In the calculation of the modeled profiles, we assumed the nominal extent of the PSF, equal to 1.0 arcsec (see Methods).

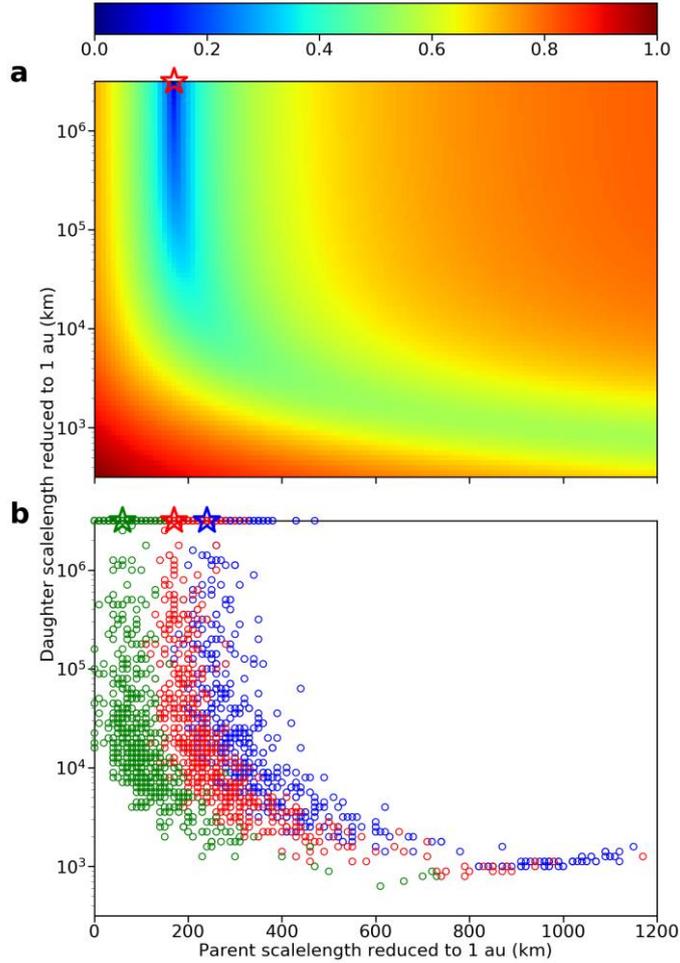

**Figure 3 | Haser scalelengths of the observed nickel emission. a,** Map of the standard deviation of the model fits from the observed profile in Fig. 2, computed for the nominal extent of the PSF, equal to 1.0 arcsec (see Methods). The linear color scale shows the quantity $\sigma_{disp} = ((\sigma - \sigma_{min})/(\sigma - \sigma_{max}))^{\gamma}$, where $\sigma$ is the mapped standard deviation of the fit and $\gamma = 0.2$ was chosen empirically. **b,** Map of the solutions resulting from 3,000 Monte Carlo simulations, constructed from the observation noise for the minimum, nominal, and maximum extents of the PSF, equal to 0.65 (blue), 1.0 (red) and 1.5 (green) arcsec, respectively (see Methods). The scatter within each PSF group shows the effect of the observation noise, whereas the spread of the groups illustrates the influence of atmospheric seeing. In both panels, the nominal solutions are indicated by the star symbols.

Finally, using the Haser model with the constrained scalelengths and the assumed gas velocity, and using the model of nickel fluorescence emission, we find from the measured fluxes of the two brightest lines (see Table 1) a nickel production rate of $0.9 \pm 0.3 \times 10^{22}$ atoms s$^{-1}$ (see Methods). Nickel is therefore a minor constituent of the coma of 2I/Borisov, with a relative abundance of 0.002% compared to OH and 0.3% compared to CN at the time of our observations (see Methods). The uncertainty of the production rate is estimated from the errors of the measured line fluxes and correlated uncertainties of the PSF and parent and daughter scalelengths (propagated from the scatter in the Monte Carlo simulation), and it additionally accounts for an (assumed) 30% cumulative uncertainty of absolute flux calibration, fluorescence efficiencies and gas velocity. As can be seen in Extended Data Fig. 2, the variation of scalelengths and PSF within the allowable regimes has a small effect on the determined nickel production rate (see Methods).

Nickel is supplied in the Universe by exploding white dwarfs and exploding massive stars, with relative contributions of 71% and 29%, respectively[33]. Solid nickel is a major constituent of interplanetary matter, and is the second most abundant chemical element incorporated in iron meteorites in the form of FeNi alloys[34]. Measurements of metal ions in the upper atmosphere after meteor showers, obtained using a rocket-borne ion mass spectrometer, showed that the abundances of metal ions—including Ni—in meteors are compatible with the abundances in chondrite meteorites[35]. Very similar metal abundances were reported from the in-situ analysis of dust around comet 1P/Halley[36]. FeNi alloys as well as iron-nickel sulfides were also present in all samples of cometary dust collected during the Stardust mission to comet 81P/Wild[37,38].

While solid-phase nickel is widespread, observation of the gaseous form of nickel (and other metals) has been limited to hot environments. Notably, emissions from neutral nickel vapor and other gaseous metals were detected in comet C/1965 S1 (Ikeya-Seki) at heliocentric distances of ~30 and ~13 solar radii[17,18]. Several years later some faint, transient emission features, consistent with nickel and silicon ions, were observed in the solar corona shortly after the possible collision of comet C/1979 Q1 (Solwind) with the Sun[39]. More recently, an iron tail was indirectly identified in comet C/2006 P1 (McNaught) at a heliocentric distance <40 solar radii[19]. Gaseous metals were also detected in hot environments around other stars. They are observed in the atmospheres of ultrahot Jupiters[40,41] at equilibrium temperatures >2,000 K, including a recent detection of atomic nickel vapor at WASP-121b[41]. Evaporation of numerous large exocomets is regarded as the most plausible explanation for the strongly variable metal absorption lines observed in the β Pic debris disk, with a measured Fe I gas temperature of ~1,300 K (ref. [20]). Disintegrating minor bodies are also thought to be the source of metal pollution in white dwarf atmospheres[42,43].

In this context, the detection of gaseous nickel at an interstellar comet traveling through the cold outskirts of the terrestrial planet region beyond 2 au from the Sun is an unexpected finding. Based on the observations of sungrazing and sunskirting comets, the equilibrium temperature needed for rocky dust to sublimate is >700 K; this is high compared to the 180 K equilibrium temperature of 2I/Borisov at the time of our observations (see Methods). Even the subsolar equilibrium temperature of 260 K—which is consistent with the maximum nucleus temperature from realistic thermal modeling[44]—is much too low. In the absence of known non-solar high-temperature heat sources, strong enough to cause dust sublimation at large heliocentric distances, and given the detected Haser scalelength of the nickel parent, unbound nickel atoms seem to originate from the photodissociation of a short-lived nickel-bearing molecule that sublimates at low temperatures or is otherwise released with major volatiles.

Until recently, atomic nickel has escaped detection in the cold comae of Solar System comets. Notably, nickel was not detected by the Rosetta spacecraft at comet 67P/Churyumov-Gerasimenko in any form, and no heavy element was observed in the gas phase, despite the smaller heliocentric distance of 1.5 au (ref. [45]). However, an independent study shows that gaseous nickel (and iron) is indeed present in native Solar System comets up to large heliocentric distances, but has been overlooked in previous studies[21]. The simultaneous identification of this species in the cold comae of both 2I/Borisov and local comets shows that they have even more in common than was previously thought, and strengthens the affinity between the unknown birthplace of 2I/Borisov and our own Solar System. However, whether analogs of Solar System minor bodies are widespread throughout the Galaxy is yet to be seen; although 2I/Borisov is remarkably familiar, the only previous known interstellar visitor 1I/'Oumuamua was quite the opposite[3-6]. New discoveries of interstellar minor bodies are expected at an increased rate after the commencement of the Vera C. Rubin Legacy Survey of Space and Time[46], which might help us to answer this question.

## Methods

**Additional details of the observations**

As part of our custom calibration plan, we secured accurate absolute flux calibration by observing, on each night, spectrophotometric standard stars HD 111980 and HD 115169. The former was observed before the comet at a higher airmass and the latter after the comet at a lower airmass. Because of the extended nature of our target and the rather limited slit length, our nightly routine also included frequent integrations of the sky background at offset positions (10.0 arcmin on the first night and 5.0 arcmin on the following two nights). On the first night we took 400-s integrations and obtained 13 individual spectra of 2I/Borisov (O) and 6 spectra of the sky background (S) in the sequence OSO, whereas on the second and third nights we took 240-s integrations, obtaining on each night 18 spectra of 2I/Borisov and 9 spectra of the sky background in the sequence OOS. All spectra were taken with the slit aligned with parallactic angle. Exposures of arc-lamps and flat-fields were also obtained as part of the X-shooter default calibration plan.

The target airmass was in the range 1.14 – 1.45 and the mean zenithal size of the seeing disk at 500 nm was equal to 0.76 arcsec in full width at half maximum (FWHM), extracted from the seeing measurements recorded at the start and end of the exposures. To determine the extent of the PSF applicable to our combined spectrum at the wavelengths of nickel emission, we scaled the recorded seeing values for airmass and wavelength using the canonical power laws with exponents of +0.6 and -0.2, respectively. The resulting effective seeing was equal to 0.92 arcsec, suggesting that the overall extent of the PSF (atmospheric and instrumental) was close to 1.0 arcsec, and certainly within the range 0.65 – 1.5 arcsec.

**Data processing**

All spectra were first processed using the ESO Reflex pipeline[47] to merge the orders, calibrate the wavelength (using arc-lamp spectra), correct for the flat-field (using flat-field integrations) and subtract inter-order background. We then cleaned the spectra of cosmic rays and other minor artefacts, filling them with the interpolated neighborhood values.

Visual inspection of the individual integrations of 2I/Borisov showed four spectra in which the comet was out of the slit and five spectra in which the comet's signal was contaminated by background objects (cross-checked in Sloan Digital Sky Survey's images). These spectra were removed from further analysis. As an auxiliary procedure, the 2D spectra of the comet and sky background were summed along the spectral dimension to reveal faint background objects, which were then completely masked.

In each 2D sky spectrum, the signal along the spatial dimension was median-combined to produce a 1D version of the spectrum. Pairs of the 1D sky spectra bracketing integrations of the comet were then linearly interpolated to the middle time of the comet spectrum, and the result of this procedure was subtracted from every row of the comet's spectrum, effectively removing the sky-emission component. The resulting spectra were then flux-calibrated in the usual manner using observed and reference[48] spectra of two spectrophotometric standard stars (see Methods section 'Additional details of the observations'). The reference spectra were converted to air wavelengths[22] and the observed spectra (natively in the air reference frame) were smoothed with a Gaussian kernel to match the spectral resolution and sampling of the reference spectra. This procedure enabled us to retrieve the extinction curve and instrumental response function for each night.

Repeating the previously described auxiliary procedure, we summed the sky-subtracted and flux-calibrated spectra of 2I/Borisov along the spectral dimension (wavelength range 3,600.0 – 5,400.0 Å) to find the brightness peak of the coma. This enabled us to vertically shift and average the individual 2D spectra. The resulting merged 2D spectrum had increased noise in the rows close to the edges of the spatial dimension; therefore, for the analysis we extracted the central part (64 pixels long; equivalent to 10.24 arcsec or 15,300 km at the comet). Subsequently, we summed this spectrum along the spatial dimension to create a 1D high signal-to-noise ratio wavelength-calibrated and flux-calibrated spectrum that was free of atmospheric emissions.

To remove the dust-continuum component, we selected two emission-free regions (4,390.0 – 4,500.0 Å and 5,200.0 – 5,320.0 Å) and least-square-fitted the spectrum of the Sun with an unrestricted linear spectral slope. For this procedure, we used a high-spectral-resolution solar irradiance atlas[49], which we converted to air wavelengths[22], Doppler-shifted to account for the heliocentric and geocentric velocities of the comet, and smoothed with a Gaussian kernel to match the spectral resolution and sampling of our data. The fit was then subtracted from the 1D spectrum of 2I/Borisov to produce the final spectrum that was used for the analysis.

**Spectral line measurements**

The line fluxes were consistently measured in 2-Å wavelength intervals (centered to a fraction of a spectral channel to maximize the recorded flux), and the noise was estimated from the nearest emission-free intervals to the left and right with a 5-Å width. The width of the flux interval was chosen as a compromise between the need to enclose the entire flux and cut out the unwanted noise. Thus, the actual signal-to-noise ratios of the detections might be higher.

**Nickel fluorescence**

To compute the fluorescence spectrum of nickel, we first solved statistical-equilibrium equations for 133 energy levels participating in 464 transitions listed in the Atomic Spectra Database of the National Institute of Standards and Technology (NIST)[50]. The equations balance the fluorescence absorption and emission, and can be written for each energy level $i$ in the form:

$$\sum_{u>i} \phi_u \mathcal{R}_{ui} + \sum_{l<i} \phi_l \mathcal{R}_{li} = \phi_i \left( \sum_{u>i} \mathcal{R}_{iu} + \sum_{l<i} \mathcal{R}_{il} \right), \tag{1}$$

where $\phi_i$ is the fractional population of the *i*-th energy level, $\mathcal{R}_{lu}$ is the pumping rate between the lower level *l* and the upper level *u*, and $\mathcal{R}_{ul}$ is the decay rate between these levels. The pumping and decay rates are defined through the Einstein coefficients for spontaneous emission $A_{ul}$, prompt emission $B_{ul}$, and absorption $B_{lu}$:

$$\mathcal{R}_{lu} = B_{lu}\mathcal{F}_\odot(\lambda) = A_{ul} \frac{w_u}{w_l} \frac{\lambda_{ul}^5}{8\pi hc^2} \mathcal{F}_\odot(\lambda), \tag{2}$$

$$\mathcal{R}_{ul} = A_{ul} + B_{ul}\mathcal{F}_\odot(\lambda) = A_{ul}\left(1 + \frac{\lambda_{ul}^5}{8\pi hc^2}\mathcal{F}_\odot(\lambda)\right), \tag{3}$$

where $h = 6.62607015 \times 10^{-34}$ J s is the Planck constant, $c = 2.99792458 \times 10^8$ m s$^{-1}$ is the speed of light, $\lambda_{ul}$ is the transition wavelength, $w_u$ and $w_l$ are the statistical weights of the upper and lower energy levels, and $\mathcal{F}_\odot(\lambda)$ is the wavelength-dependent solar energy flux density (energy per unit time, unit surface area, and unit wavelength). $\mathcal{F}_\odot(\lambda)$ can be calculated from the solar irradiance spectrum (i.e. energy flux density at 1 au from the Sun) by scaling with the inverse square of heliocentric distance. Note that the right-hand sides of the last two equations were obtained by substituting the known relations between Einstein coefficients. The Einstein coefficient for spontaneous emission $A_{ul}$ was taken from the NIST data. From the provided error flags we conclude that the errors of $A_{ul}$ of the nine detected lines of nickel range between ≤18% and ≤25%. Moreover, in the computation of the absorption and prompt emission rates we used solar irradiance data from two sources: for the vacuum wavelength range 2,990.0 – 10,000.0 Å we used the high-resolution solar atlas[49] obtained with the spectral resolution of 500,000 and natively available in the vacuum reference frame (the same that we used for the dust continuum subtraction; see earlier in Methods), and for the range 2,000.6 – 2,990.0 Å we used 0.1-Å data[51], which we converted to vacuum wavelengths[22]. This composite solar irradiance spectrum was then scaled by the inverse square of heliocentric distance, and Doppler-shifted in accordance with the comet's heliocentric velocity to account for the Swings effect[28]. The resulting set of equations is linearly dependent, but the dependence is removed after replacing one (any) of the equations with a normalization condition:

$$\sum_i \phi_i = 1. \tag{4}$$

The set was solved with the LU decomposition method. As a result of this computation, we obtained fractional level populations $\phi_i$ for each of the 133 considered energy levels, and then obtained the fluorescence efficiencies $g_{ul}$ for the 464 considered transitions using:

$$g_{ul} = \frac{hc}{\lambda_{ul}} \phi_u A_{ul}. \tag{5}$$

The fluorescence efficiency is conventionally reduced to 1 au from the Sun, obtained by scaling with the inverse square of heliocentric distance.

The model was tested against the spectrum of comet C/1965 S1 (Ikeya-Seki) and provided an excellent fit to the spectrum taken at a heliocentric distance of ~30 solar radii[17]. A fit to the spectrum taken at ~13 solar radii[18], although still reasonable, is not as good, but consistent with saturation effects of the photographic plate, and possibly affected by additional excitation and de-population mechanisms that operate at the boundary of the solar corona.

**Haser model**

The density distribution of neutral species in the cometary coma is conventionally described by the Haser model[29,30]. We used the standard two-generation version, in which isotropically ejected first-generation 'parent' species travel radially outwards and progressively decay, giving birth to second-generation 'daughter' species that continue along the same lines at the same speed, also progressively decaying. In this simple scenario, the number of daughter species $dn_d$ contained in an infinitesimal volume $dV$ is:

$$dn_d(r) = \frac{Q}{4\pi r^2 v} \frac{\rho_d}{\rho_p - \rho_d} \left(e^{-r/\rho_p} - e^{-r/\rho_d}\right) dV, \tag{6}$$

where $r$ is the nucleocentric distance, $Q$ and $v$ are the production rate and speed of the species (the same for both generations in the assumed scenario), and $\rho_p$ and $\rho_d$ are the parent and daughter scalelengths; their corresponding lifetimes are $\tau = \rho/v$. For equal $\rho_p$ and $\rho_d$ the above equation becomes:

$$dn_d(r) = \frac{Q}{4\pi r^2 v} \frac{r}{\rho} \left(e^{-r/\rho}\right) dV, \tag{7}$$

where $\rho = \rho_p = \rho_d$ is the common scalelength. Integration of $dn_d$ over infinite volume gives the total number of daughter species in the coma:

$$n_d = \frac{Q \rho_d}{v}. \tag{8}$$

Note that the shape of the daughter density distribution depends solely on the two scalelengths and can be conveniently traced through the fraction of daughter species $df_d$ contained in an infinitesimal volume $dV$:

$$df_d(r) = \frac{dn_d(r)}{n_d} = \frac{1}{4\pi r^2} \frac{1}{\rho_p - \rho_d} \left(e^{-r/\rho_p} - e^{-r/\rho_d}\right) dV, \tag{9}$$

which becomes:

$$\mathrm{d}f_d(r) = \frac{1}{4\pi r^2}\frac{r}{\rho^2}\left(e^{-r/\rho}\right)\mathrm{d}V, \tag{10}$$

for equal $\rho_p$ and $\rho_d$ scalelengths. Integration of $\mathrm{d}f_d$ over infinite volume obviously returns the normalization condition $f_d = 1$. It can be readily seen that the fraction of species $\mathrm{d}f_d$ is invariant under the interchange of $\rho_p$ and $\rho_d$, making it impossible to attribute the scalelengths to the generations based on the shape of the density profile alone. The model becomes indefinite if both scalelengths are equal to zero; however, if only one scalelength is equal to zero, it becomes effectively reduced to the simpler single-generation version.

Although the Haser model ignores the additional isotropic velocity gained by the daughter species at creation, it can still produce realistic daughter density profiles and production rates, albeit with altered scalelengths and speed to compensate for the missing parameter[30,52]. Likewise, the implicated equality of the parent and daughter production rates is often poorly realized, as most parent species have multiple decay branches and most daughter species can be created via multiple avenues. However, the two production rates may still be calculated from one another if the parent-daughter photochemical path is well characterized in terms of the parent decay ratio and daughter origin ratio. In the calculation of nickel production rate we assumed that nickel originated from a single parent, but refrained from associating this production rate with the parent due to the unknown branching ratios of the parent.

The lifetime of the species changes with the square of heliocentric distance and it is also dependent on the level of solar activity. It is normally given at a standard heliocentric distance of 1 au to facilitate comparisons between species and comets. The scalelength changes in the same way if the species velocity is set constant. In this work we used the standard scaling with the square of heliocentric distance.

**Haser scalelengths**

To retrieve the two Haser scalelengths from our data, we generated a set of 9,801 synthetic Haser profiles along the slit length with different combinations of scalelengths and compared these profiles against the observed spatial profile (Fig. 2) in search of the best match. Each synthetic profile was calculated on a grid emulating the 1.3 × 10.24 arcsec (1,950 × 15,300 km) trimmed slit and consisting of 45 × 320 cells, oversampling the actual number of CCD pixels along the slit by a factor of 5. First, we integrated the fractions of the daughter species $\mathrm{d}f_d$ along the line of sight using our in-house adaptive step-size integrator with error control. Next, the integrated fractions were convolved with the PSF approximated by a symmetric two-dimensional Gaussian with FWHM equal to 1.0 arcsec (see Methods section 'Additional details of the observations'). Finally, we binned the grid cells using a 45 × 5 bin size, which resulted in a one-dimensional synthetic profile along the slit sampled at 64 equally spaced points, consistent with the observed spatial profile. Note that the fractions $\mathrm{d}f_d$ were consistently calculated with the comet nucleus offset from the slit center by -0.083 arcsec (-0.52 pixels) along the slit length, consistent with the offset of the observed profile

(determined from the best match of a single-generation infinite-lifetime density profile with an unrestricted offset along the slit). The entire set of synthetic profiles was based on 121 parent scalelengths equally spaced along a linear scale between 0 and 1,200 km, and 81 daughter scalelengths equally spaced along a logarithmic (base of 10) scale between $10^{2.5}$ and $10^{6.5}$ km, both at 1 au from the Sun. Every profile from the set was individually scaled to best match the observed profile by minimizing the sum of squared residuals. As a result of this procedure, we could construct a map of standard deviation (Fig. 3a) and therein find the best solution.

**Monte Carlo simulations**

We used a Monte Carlo method to simulate the effect of the observation noise and seeing uncertainty on the determined Haser scalelengths and the production rate. The simulations were generated and analyzed in three groups corresponding to the PSF FWHMs of 0.65, 1.0, and 1.5 arcsec, which we considered to be, respectively, the minimum, the nominal, and the maximum values applicable to our combined spectrum of nickel (see Methods section 'Additional details of the observations'). Each group consisted of 1,000 clones of the observed spatial profile along the slit length. The clones were generated from the best-matching synthetic Haser profile (with a group-specific PSF) by superimposing random realizations of noise. We assumed normally-distributed noise with a constant standard deviation calculated as the root-mean-square deviation of the best-matching profile from the observed one. As with the original search for the Haser scalelengths, the clones from each group were linked with the best-matching combinations of scalelengths identified in a set of 9,801 synthetic Haser profiles with a compatible PSF, and used for the determination of a corresponding production rate. For details of the Haser profile computation see Methods section 'Haser scalelengths'.

**Production rate**

The production rate can be easily deduced from the observed emission. Assuming an optically thin regime, the measured energy flux $F$ is simply the energy rate of a single emitter $g_{ul}$, known as the fluorescence efficiency or 'g-factor' (see Methods section 'Nickel fluorescence') multiplied by the number of emitters visible in the slit and divided by the area of a spherical surface at the distance of the observer $\Delta$. Thus,

$$F = \frac{n^* g_{ul}}{4\pi \Delta^2}, \qquad (11)$$

where $n^* = n f^*$, and the asterisk indicates that the quantities were calculated upon integration over the slit, to distinguish from the corresponding quantities calculated with infinite integration limits, introduced earlier in Methods section 'Haser model'. By further assuming the Haser density distribution and substituting Equation (8), we obtain:

$$Q = \frac{4\pi \Delta^2 v}{\rho f^* g_{ul}} F. \tag{12}$$

Note that the last two equations hold for species of any generation.

**Reference production rates of OH and CN**

To facilitate a comparison of the production rate of nickel with common cometary species in 2I/Borisov, we calculated the production rates of OH and CN, both robustly detected in our data (see Extended Data Fig. 1). The flux of OH was measured to be $4.6 \pm 0.9 \times 10^{-15}$ erg s$^{-1}$ cm$^{-2}$ (integrated between 3,070.0 and 3,110.0 Å) and the flux of CN was equal to $3.73 \pm 0.02 \times 10^{-15}$ erg s$^{-1}$ cm$^{-2}$ (integrated between 3,850.0 and 3,885.0 Å). These fluxes imply OH and CN production rates of $3.6 \times 10^{26}$ molec s$^{-1}$ and $2.5 \times 10^{24}$ molec s$^{-1}$, respectively, calculated using the standard two-generation Haser model (see Methods section 'Haser model') with the canonical parameters[31] and using the fluorescence efficiencies of $5.0 \times 10^{-15}$ erg s$^{-1}$ and $3.65 \times 10^{-13}$ erg s$^{-1}$, respectively, both applicable to our observing geometry[53,54]. Note that in this calculation we used a higher species' speed of 1 km s$^{-1}$ (compared to 0.5 km s$^{-1}$ used for nickel); this reflects the excess speed that OH and CN gain at creation (see Methods section 'Haser model'). The excess speed of nickel is unfortunately unknown, and as such it was neglected. A detailed study of these and other common species in the coma of 2I/Borisov, based on the X-shooter data, will be presented in a separate work.

**Equilibrium temperature**

The equilibrium temperature is the temperature that equilibrates the incoming and outgoing energy. The average equilibrium temperature $T_{av}$ of a rotating spherical body orbiting the Sun is given by:

$$T_{av} = \left(\frac{L_\odot (1-A)}{16\pi\sigma R^2}\right)^{1/4}, \tag{13}$$

where $\sigma = 5.670374419 \times 10^{-8}$ W m$^{-2}$ K$^{-4}$ is the Stefan-Boltzmann constant, $L_\odot = 3.828 \times 10^{26}$ W is the solar luminosity, $A$ is the Bond albedo, and $R$ is the heliocentric distance. The maximum equilibrium temperature is reached at the subsolar point. In the extreme case of the rotation axis directed to the Sun (or a non-rotating body), the subsolar equilibrium temperature $T_{ss}$ becomes:

$$T_{ss} = \sqrt{2} T_{av}. \tag{14}$$

In the calculation of the equilibrium temperatures of 2I/Borisov, we implicitly assumed a typical cometary albedo of $A = 0.04$. Note that the result is weakly dependent on the albedo as long as $A \ll 1$.

## Supplementary references

## Data availability

The X-shooter raw data are available in the ESO archive at https://archive.eso.org. Source data are provided with this paper.

## Code availability

The EsoReflex pipeline is available from the ESO website at: https://www.eso.org/sci/software/esoreflex/. All custom codes are direct implementations of standard methods and techniques, described in detail in the Methods section.

## Competing financial interests

The authors declare no competing interests.

## Acknowledgements

We thank Krzysztof Rusek for help with proposal writing, Milena Ratajczak and Mariusz Gromadzki for introducing us to X-shooter data reduction, and Paweł Kozyra for discussion on nickel-bearing molecules. This work is based on observations collected at the European Southern Observatory under ESO programme 0104.C-0933(B). We thank the ESO staff for support. We are also grateful for support from the National Science Centre of Poland through ETIUDA scholarship no. 2020/36/T/ST9/00596 to P.G. and SONATA BIS grant no. 2016/22/E/ST9/00109 to M.D., and we acknowledge support from the Polish Ministry of Science and Higher Education through grant no. DIR/WK/2018/12.

## Author contributions

P.G. and M.D. wrote the telescope time proposal, searched for the origin of the detected spectral lines, and wrote the paper. P.G. prepared the observations, reduced and calibrated the data, identified the emitting species, and measured the spectral lines. M.D. created the fluorescence model, retrieved the scalelengths, and calculated the production rate.

## Author information

Correspondence and requests for materials should be addressed to P.G. (piotr.guzik@doctoral.uj.edu.pl) or M.D. (drahus@oa.uj.edu.pl).

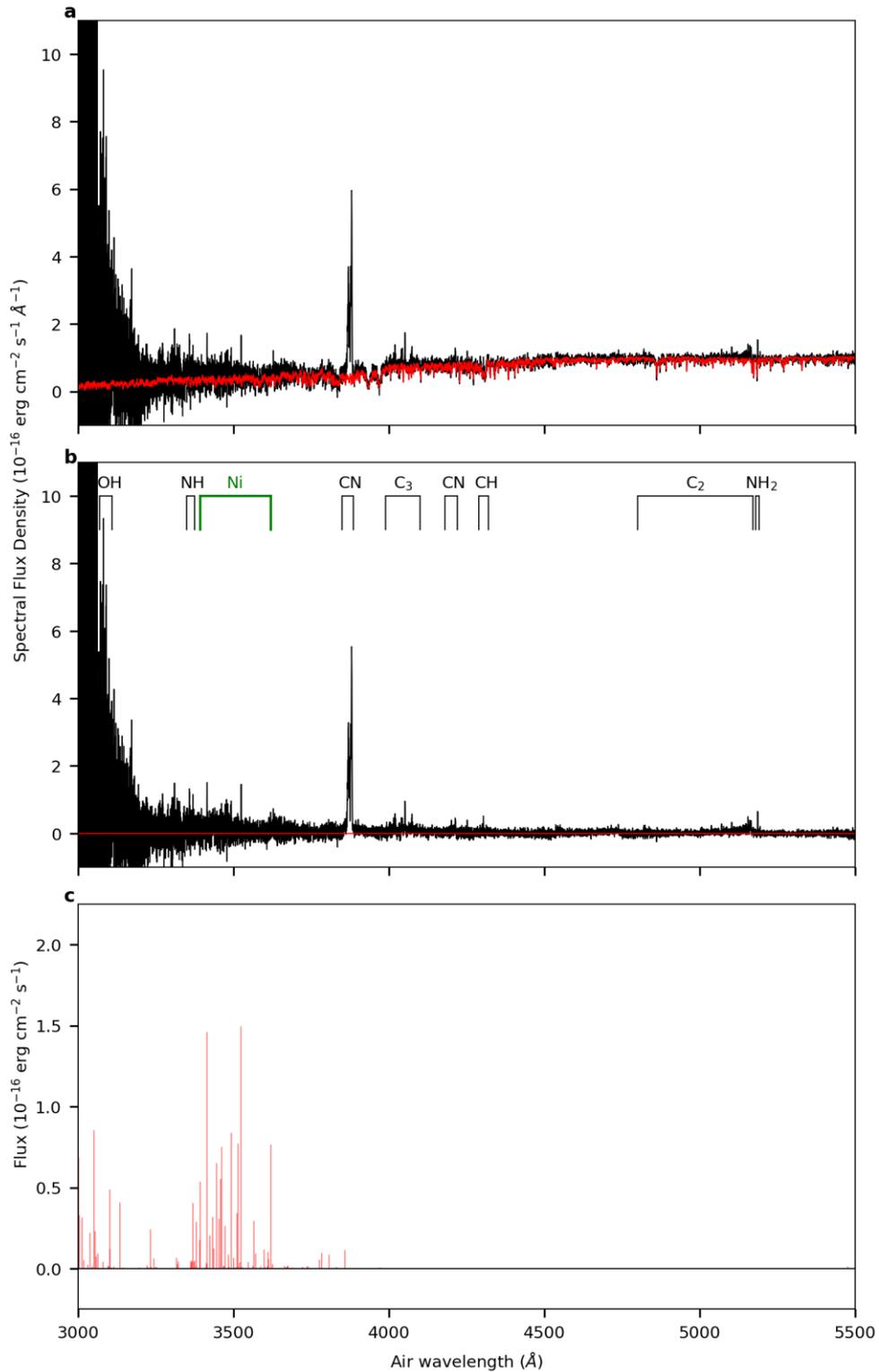

**Extended Data Figure 1 | Complete spectrum of comet 2I/Borisov from X-shooter UVB arm. a**, Flux-calibrated spectrum with fitted dust continuum (see Methods). **b**, Same as **a** but with the dust-continuum component removed. Major emission features are labeled. **c**, Modeled spectrum of nickel fluorescence emission (see Methods), converted to air wavelengths[22], and scaled to best match the two brightest lines.

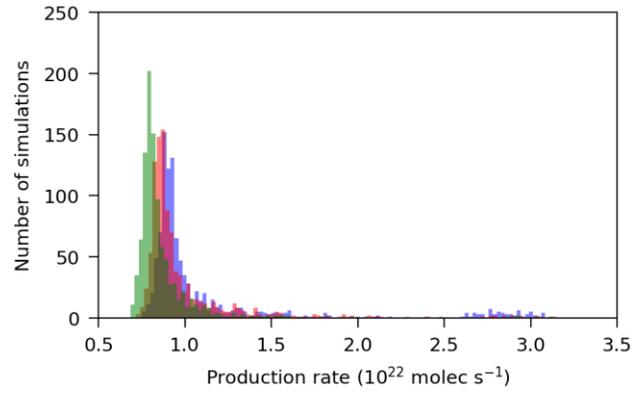

**Extended Data Figure 2 | Distribution of Monte Carlo-simulated production rates.** The distribution was constructed from the production rates corresponding to the results of the Monte Carlo simulation in Fig. 3b (see Methods). Results are presented in three groups according to the assumed PSF equal to 0.65 (blue), 1.0 (red) and 1.5 (green) arcsec.